\documentclass[%
 reprint,
superscriptaddress,
 amsmath,amssymb,
 aps,
prl
]{revtex4-1}

\usepackage{amssymb}
\usepackage{fullpage}
\usepackage{graphicx}
\usepackage{bibunits}
\usepackage{wrapfig}
\usepackage{floatflt}
\usepackage{color}
\usepackage{xspace}
\usepackage{dsfont}
\usepackage[official,right]{eurosym}
\usepackage[top=0.8in, bottom=0.75in, left=0.75in, right=0.75in]{geometry}

\usepackage{graphicx}
\usepackage{dcolumn}
\usepackage{bm}
\usepackage{hyperref}

\usepackage{pdfpages}
\makeatletter
\AtBeginDocument{\let\LS@rot\@undefined}
\makeatother

\usepackage{multirow}

\newcommand{\KCUC} {K$_2$CuSO$_4$Cl$_2$\xspace}

\newcommand{\BPCB} {(C$_5$H$_{12}$N)$_2$CuBr$_4$\xspace}
\newcommand{\CuPZN}{Cu(pz)(NO$_3$)$_2$\xspace}

\newcommand{\be}{\begin{equation} }
\newcommand{\ee}{\end{equation} }
\newcommand{\bea}{\begin{eqnarray} }
\newcommand{\eea}{\end{eqnarray} }

\newcommand{\mb}[1]{\mathbf{#1}}
\newcommand{\mr}[1]{\mathrm{#1}}

\def\XXint#1#2#3{{\setbox0=\hbox{$#1{#2#3}{\int}$}
     \vcenter{\hbox{$#2#3$}}\kern-.5\wd0}}

\begin{document}

\title{$z=2$ Quantum Critical Dynamics in a Spin Ladder}

\author{D. Blosser}
\email{dblosser@phys.ethz.ch}
\affiliation{Laboratory for Solid State Physics, ETH Z\"urich, 8093 Z\"urich, Switzerland}

\author{V. K. Bhartiya}
\affiliation{Laboratory for Solid State Physics, ETH Z\"urich, 8093 Z\"urich, Switzerland}

\author{D. J. Voneshen}
\affiliation{ISIS Facility, Rutherford Appleton Laboratory, Chilton, Didcot, Oxon OX11 0QX, United Kingdom}

\author{A. Zheludev}
\email{zhelud@ethz.ch}
\homepage{http://www.neutron.ethz.ch/}
\affiliation{Laboratory for Solid State Physics, ETH Z\"urich, 8093 Z\"urich, Switzerland}

\date{\today}

\begin{abstract}
By means of inelastic neutron scattering we investigate finite temperature dynamics in the quantum spin ladder compound \BPCB (BPCB) near the magnetic field induced quantum critical point with dynamical exponent $z=2$. We observe universal finite-temperature scaling of the transverse local dynamic structure factor in spectacular quantitative agreement with long-standing theoretical predictions. At the same time, already at rather low temperatures, we observe strong non-universal longitudinal fluctuations. To separate the two, we make use of an intrinsic leg-exchange symmetry of the spin ladder. Complementary measurements of specific heat also reveal striking scaling behavior near the quantum critical point. 
\end{abstract}


\maketitle

Scaling is one of the cornerstone concepts in modern physics and plays a key role in understanding critical phenomena, particularly quantum phase transitions (QPTs) \cite{SachdevBook1999,Vojta2003,SachdevKeimer2011,Sachdev2008_MagnetismAndCriticality}. Near criticality all physical properties are described by a set of critical exponents and scaling functions. These are {\em universal} and only depend on the symmetry and dimensionality of the problem (the universality class of the transition), but not on the microscopic details of the Hamiltonian \cite{Stanley1971introduction}. Unfortunately, the actual computation of critical exponents and particularly the scaling functions is rarely possible analytically \cite{HEStanley1999}.  When it is though, these exact results can be quantitatively applied to real word phase transitions, without knowing almost anything about the microscopic details of the experimental system.

One notable analytically solvable case are field-induced quantum phase transitions in one-dimensional ($d=1$) quantum Heisenberg antiferromagnets (HAFs) between their various gapped and gapless phases. These transitions are understood in terms of a ``condensation'' of bosonic quasiparticles with a hard-core repulsion, where the magnetic field plays the role of a chemical potential \cite{Affleck1991,Sachdev1994}. The dynamical critical exponent is $z=2$. All other critical exponents \cite{Affleck1991,Sachdev1994ConservedCharges,Sachdev1994} and scaling functions are also known excatly \cite{Sachdev1994, Korepin1990,Girardeau1960}. A series of recent experiments on the $S=1/2$ chain compound \CuPZN (pz denotes pyrazine) \cite{BreunigLorenz2017, JeongRonnow2015,Kono2015,Rohrkamp2010,Kuhne2009,Kuhne2011} and the spin ladder system \BPCB (BPCB) \cite{Watson2001,Lorenz2008} have provided a spectacular verification of the predicted scaling behavior of thermodynamic quantities and quasistatic local fluctuations. However, the scaling functions for {\em spatial and temporal correlations} at this transition are also fully universal. Calculating them exactly is a formidable task, but has been accomplished  \cite{Korepin1990, Korepin1997Book, Barthel2012, PanfilJSCaux2014, Blosser2017}. Can they be also measured experimentally?

One recent attempt was made in inelastic neutron scattering experiments on the  spin chain compound \KCUC near magnetic saturation \cite{Blosser2017}. The measurement of quantum critical correlations largely failed due to an unexpected problem: for spin chains, the relevant scattering from critical spin fluctuations transverse to the applied magnetic field overlaps with that from strong non-critical longitudinal fluctuations. In the present work we overcome this seemingly insurmountable hurdle by performing similar measurements on a spin ladder material, and exploiting the ladder's intrinsic rung-exchange symmetry to separate the two types of scattering. We measure the scaling function for the local dynamic structure factor over one and a half decades in $\hbar \omega/k_B T$ and find a good agreement with exact results for hard core bosons. We also measure the scaling of specific heat, and discuss some peculiar non-universal features of the excitation spectrum.

\begin{figure}
\includegraphics{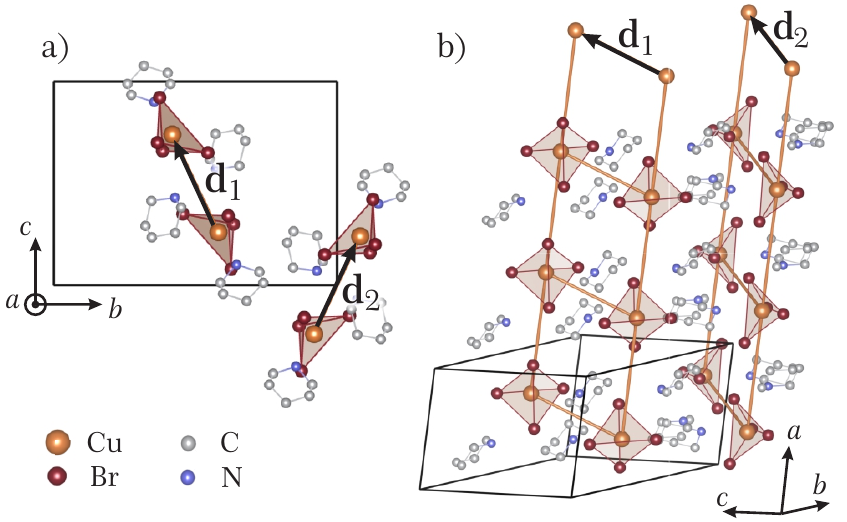}
\caption{\label{fig:structure}Schematic of the crystal structure of BPCB showing the ladders formed by the magnetic Cu$^{2+}$ cations and linking Br$^{-}$ anions. Hydrogen atoms of the organic ligands are not shown. The vectors $\mb{d}_{1,2}$  define the rung orientation of the two inequivalent ladders in each unit cell.}
\end{figure}

Our target compound is the well known strong-rung HAF $S=1/2$ ladder system \BPCB (BPCB) \cite{Patyal1990}. The spin ladders are formed by magnetic Cu$^{2+}$ cations linked by super-exchange bridges via Br$^-$ anions \cite{Savici2009}. They run along the $a$ axis of the monoclinic crystal structure (P2$_1$/c, $a=8.49$, $b=17.22$, $c=12.38$~\AA, $\beta=99.3^{\circ}$ \cite{Patyal1990}), and are well separated by non-magnetic organic piperidinium molecules (Fig. \ref{fig:structure}). The ground state is a quantum spin singlet  \cite{Ruegg2008BPCBThermodynamics} with a gap $\Delta\approx 0.8$~meV in the magnetic excitation spectrum \cite{Savici2009}. In an applied magnetic field the degeneracy of the triplet band is lifted by the Zeeman effect \cite{ThielemannRuegg2009}. The present work is focused on the $z=2$ quantum phase transition that occurs at a critical magnetic field $H_c=6.66(6)$~T applied along the crystallographic $b$ axis 
\footnote{The value of the critical field was determined from inelastic neutron scattering data as described in the supplement \cite{SM}.}, 
at which the gap for the lowest magnon branch closes and the ladder starts to get magnetized \cite{ThielemannRuegg2009, Watson2001}. For all our measurements we use fully deuterated single crystal BPCB samples grown from a saturated ethanol solution by slow evaporation.

\begin{figure}
\includegraphics{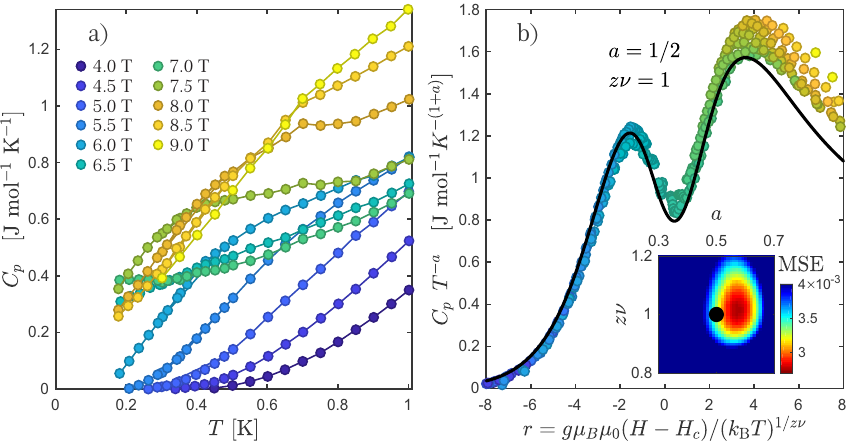}
\caption{\label{fig:SpecificHeat}a) Representative constant-field measurements of the magnetic contribution to specific heat in BPCB for a magnetic field along the $b$ axis. b) Symbols: measured magnetic specific heat for $0.17\leq T \leq 0.5$ plotted in scaled variables with the scaling exponents $a=1/2$ and $z\nu=1$. The solid line is the exact scaling function for free fermions in one dimension plotted with no adjustable parameters. Inset: False color plot of the empirical measure of the quality of the data collapse vs. the critical exponents as described in the text. Optimal scaling is found for $a=0.57(10)$ and $b=1.01(10)$.}
\end{figure}

As a first step, we verify that the thermodynamics of the transition is quantitatively consistent with exact results for the $d=1$, $z=2$ quantum critical point. The scaling form of specific heat is given by $C_p T^{-a} = \mathcal{E}(r)$, where $r=g\mu_\mathrm{B}\mu_0(H-H_c)(k_\mathrm{B}T)^{-z\nu}$ is the scaled magnetic field. The spin stiffness $m=\hbar^2/J_\mr{Min}$, $J_\mr{Min}/k_B=5.07(10)$~K is directly determined from the measured magnon dispersion (see supplement, \cite{SM}). Due to universality, the scaling exponents
$a=1/2$ and $z\nu=1$ and the scaling function
\begin{equation} 
\mathcal{E}(r)=\frac{N_A k_\mathrm{B} }{\pi\hbar} {\sqrt{2 m k_\mathrm{B}}}\int_0^\infty dx \frac{e^{x^2-r}(x^2-r)^2}{(e^{x^2-r}+1)^2}. \label{e}
\end{equation}
are exactly as for a gas of non-interacting Fermions with a quadratic diespersion \cite{Sachdev1994,Girardeau1960}.
Note that Eq.~\ref{e} contains {\em no arbitrary scaling factors} \cite{Sachdev1994}. It is plotted in Fig.~\ref{fig:SpecificHeat}(b) as a solid line.

To compare this prediction to experiment, we carried out relaxation measurements of the magnetic specific heat of BPCB in a wide range of magnetic fields and temperatures around the transition. The data were collected on a 0.72(4)~mg single crystal using a Quantum Design PPMS equipped with a $^3$He-$^4$He dilution refrigerator insert. A calculated small nuclear spin contribution and a phonon contribution extrapolated from high temperatures were subtracted from the measured data. Typical data are shown in Fig. \ref{fig:SpecificHeat}(a). In agreement with expectation, at $H<H_c$ where the spectrum is gapped, the heat capacity shows characteristic activation behavior. At $H>H_c$, on the other hand, the observed temperature dependence is linear, characteristic of the gapless Tomogana-Luttinger liquid (TLL) regime \cite{Giamarchi2004Book}.
To independently determine whether the data obey scaling, we adopted the approach described in Ref.~\cite{JeongRonnow2015}.
For every pair of critical exponents $(a,z\nu)$ the scaled specific heat $C_p T^{-a}$ was fit to a 5\textsuperscript{th} degree polynomial of the scaled field $r=g\mu_\mathrm{B}\mu_0(H-H_c)(k_\mathrm{B}T)^{-z\nu}$. The mean squared error (MSE) of the fit then serves as an empirical measure for the quality of the data collapse. Only data close to the critical point, with $0.17\leq T \leq 0.5$~K and $|r|\le 4$ were included. The resulting MSE landscape is plotted vs. $a$ and $z\nu$ in the inset in Fig. \ref{fig:SpecificHeat}(b). Optimal scaling is found for $a=0.57(10)$ and $b=1.01(10)$, in agreement with the theoretical values $a=1/2$ and $z\nu=1$.
In fact, the latter produce an excellent data collapse, as plotted in symbols in Fig.~\ref{fig:SpecificHeat}(b), where all data points in the range $0.17\leq T \leq 0.5$~K were considered. A spectacular agreement with the exact theoretical scaling function is obtained with {\em no adjustable parameters}, not even an overall scale factors. This validates BPCB and the corresponding field-induced transition as a suitable realization of the $d=1$, $z=2$ QCP. At $T<0.17$~K small 3D interactions are relevant \cite{Klanjsec2008NMROrdering,Thielemann3DOrdering2009}, whilst for $T>0.5$~K we also begin to observe deviations from scaling.

\begin{figure*}
\includegraphics[ width=18cm]{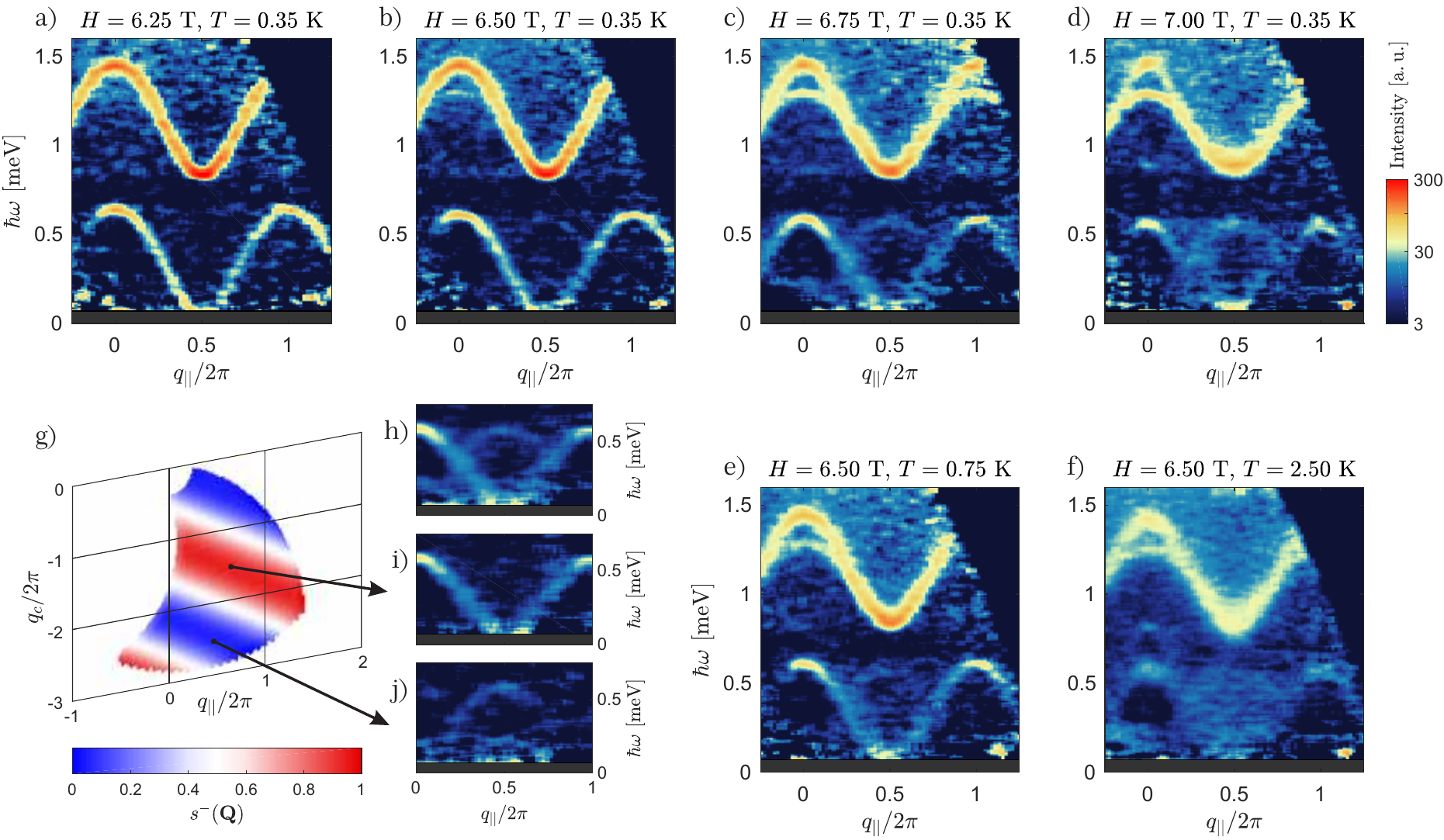}
\caption{\label{fig:NeutronSpectra}Inelastic neutron scattering spectra as measured in the spin ladder compound BPCB near the critical field $H_c=6.66(6)$~T. In the plotted range of energy transfer, the lower two of the three Zeeman-split triplet branches are visible. 
(a-d) Low temperature ($T=0.35$~K) evolution of the spectrum across the critical field. 
(b,e,f) Temperature evolution of the spectrum at $H=6.5$~T. 
(g) Antisymmetric rung structure factor $s^-(\mb{Q})$ in the covered range of reciprocal space. (h) Low energy part of the $H=6.75$~T spectrum. (i,j) Anti-symmetric and symmetric structure factor contributions determined for the same data set as described in the text. }
\end{figure*}

The main purpose of this work is a direct measurement of the scaling function for the {\em dynamic spin structure factor}. For a spin ladder, the total spin structure factor (spatial and temporal Fourier transform of the spin correlation function) $\mathcal{S}^{\alpha\alpha}(\mathbf{Q},\omega)$ can be decomposed into its symmetric and antisymmetric parts, $\mathcal{S}_+^{\alpha\alpha}(q_\|,\omega)$ and $\mathcal{S}_-^{\alpha\alpha}(q_\|,\omega)$, respectively \cite{Bouillot2011, Schmidiger2013PRB}. Here $q_\|=\mathbf{Q}\cdot\mathbf{a}$ is the wave vector transfer along the leg $\mathbf{a}$ of the spin ladder, which for BPCB coincides with the crystallographic $a$ axis. 
These two structure factors  represent correlations between the sums and differences of the two spins on each ladder rung, correspondingly. In the transition at hand, due to antiferromagnetic interactions on the ladder rungs in BPCB, it is the {\em antisymmetric} fluctuations that become critical. The dynamic structure factor is a tensor quantity, but it is the spin components that are {\em transverse} to the applied field that become critical. Thus, in regard to universal critical dynamics, the quantity of interest is $\mathcal{S}_-^{\bot}(q_\|,\omega)$. Here, we focus on the $q_\|$-integrated local dynamic structure factor. 
\begin{equation}
\mathcal{S}_-^{\perp}(\omega) = \int \mathcal{S}_-^{\bot}(q_\|,\omega)  d{q_\|}.
\end{equation}
What we seek to experimentally verify is that it follows the predicted scaling form \cite{Sachdev1994}
\begin{equation}
\mathcal{S}_-^{\perp}(\omega) = T^{-b} \Phi\left( \frac{\hbar\omega}{k_B T} \right), \qquad b=1/2, \label{phi}
\end{equation}
with a {\em completely universal} scaling function $\Phi(x)$ that is known exactly \cite{Korepin1990, Korepin1997Book, Sachdev1994, Blosser2017, Barthel2012}.

For BPCB we measured the dynamic structure factor in inelastic neutron scattering experiments \cite{DataISIS}. For this we employed 4 fully deuterated single crystal samples of total mass 2.07~g, co-aligned to better than $1^\circ$ effective mosaic spread. The measurements were carried out at the LET cold neutron time of flight spectrometer \cite{Bewley2011} at the ISIS facility, using neutrons of a fixed $2.2$~meV incident energy. The sample was mounted on a $^3$He-$^4$He dilution refrigerator in a 9~T cryomagnet, with the field applied vertical along the crystallographic $b$ direction ($z$ axis in our notation). Since all experimental scattering vectors lie close to the horizontal $(x,y)$ plane, for unpolarized neutrons, to a good approximation, the measured scattered intensity is given by
\begin{align} \label{eq:StructureFactor}
\frac{d\sigma}{d\Omega d\omega} \propto \, & s^+(\mb{Q})\left[ \mathcal{S}_+^{zz}(q_\|,\omega)+ \mathcal{S}_+^{\bot}(q_\|,\omega)  \right] \\ \notag
&+  s^-(\mb{Q})\left[ \mathcal{S}_-^{zz}(q_\|,\omega)+ \mathcal{S}_-^{\bot}(q_\|,\omega)  \right]. \notag
\end{align}
Here the rung structure factors are given by the spatial orientation of the rungs of the ladders \cite{Bouillot2011,Schmidiger2013PRB}
\begin{equation}
4 s^{\pm}(\mathbf{Q}) = 2 \pm \cos(\mb{Q}\cdot \mb{d}_1) \pm \cos(\mb{Q}\cdot \mb{d}_2),\label{sf}
\end{equation}
where $\mb{d}_{1,2}$ are the rung vectors of the two inequivalent ladders in BPCB as shown in Fig.~\ref{fig:structure}
\footnote{The rung vectors are $\mb{d}_{1,2}=[0.3904,\pm0.1598,0.4842]$ in relative lattice units for the two inequivalent ladders per unit cell in BPCB \cite{ThesisThielemann}.}.
An overview of our neutron scattering data collected near the critical field is given by the false color intensity plots in Figs. \ref{fig:NeutronSpectra}a-f).  These intensities were  integrated along the non-dispersive $b^*$ and $c^*$ directions, and therefore contain scattering from both symmetric and antisymmetric excitations. The background for all data shown in this work was collected at zero applied field at base temperature, where we assumed there to be no magnetic scattering other than in a single resolution-limited magnon band between 0.8 and 1.5~meV energy transfer \cite{Savici2009}.

The three Zeeman-split magnon branches are clearly visible in our experiments (the upper one is beyond the range of the plots shown). The critical fluctuations that are of primary interest here are descendants of the lower branch that undergoes softening. From the data shown it is obvious that  at $H>H_c$ (Fig.~\ref{fig:NeutronSpectra}c and d), but also at the critical field at elevated temperatures (Fig.~\ref{fig:NeutronSpectra}e and f) this part of the spectrum additionally contains a vague ``inverted'' band of excitations. This feature closely resembles what is observed in Heisenberg $S=1/2$ spin chains near the field of saturation \cite{Blosser2017}. A direct connection is established by the spin-ladder to spin-chain mapping described in Ref.~\cite{Bouillot2011}, and identifies the additional signal as due to non-critical longitudinal spin fluctuations \cite{Blosser2017}. As in the case of spin chains, these longitudinal excitations rapidly gain spectral weight as the ladder is magnetized (be it by increasing the magnetic field beyond $H_c$ or or by thermally populating the low-energy magnon band). Eventually they become seemingly impossible to separate from the transverse universal spin fluctuations that we are interested in.

This is where the advantage of a spin ladder over a spin chain comes into play: we can distinguish out the critical fluctuations by their leg-exchange parity, rather than by their polarization. Indeed, non-universal low energy fluctuations are due to scattering within the lower magnon band, conserve the number of magnons and therefore lie exclusively in the symmetric channel \cite{Bouillot2011}. A useful feature specific to the crystal structure of BPCB is the large contract between the led-odd and leg-even  structure factors (see Fig.~\ref{fig:NeutronSpectra}g.)  This allows us to separate the corresponding components as described in the Supplement. As an example of this analysis, Fig. ~\ref{fig:NeutronSpectra}(h) shows the low energy part of the spectrum collected at $H=6.75$~T, $T=0.35$~K, as projected over all of reciprocal space, with the two contributions overlapping. The separated symmetric and anti-symmetric contributions are shown in Fig.~\ref{fig:NeutronSpectra}(i) and (j), respectively.

\begin{figure} [t]
\includegraphics[width=8.6cm]{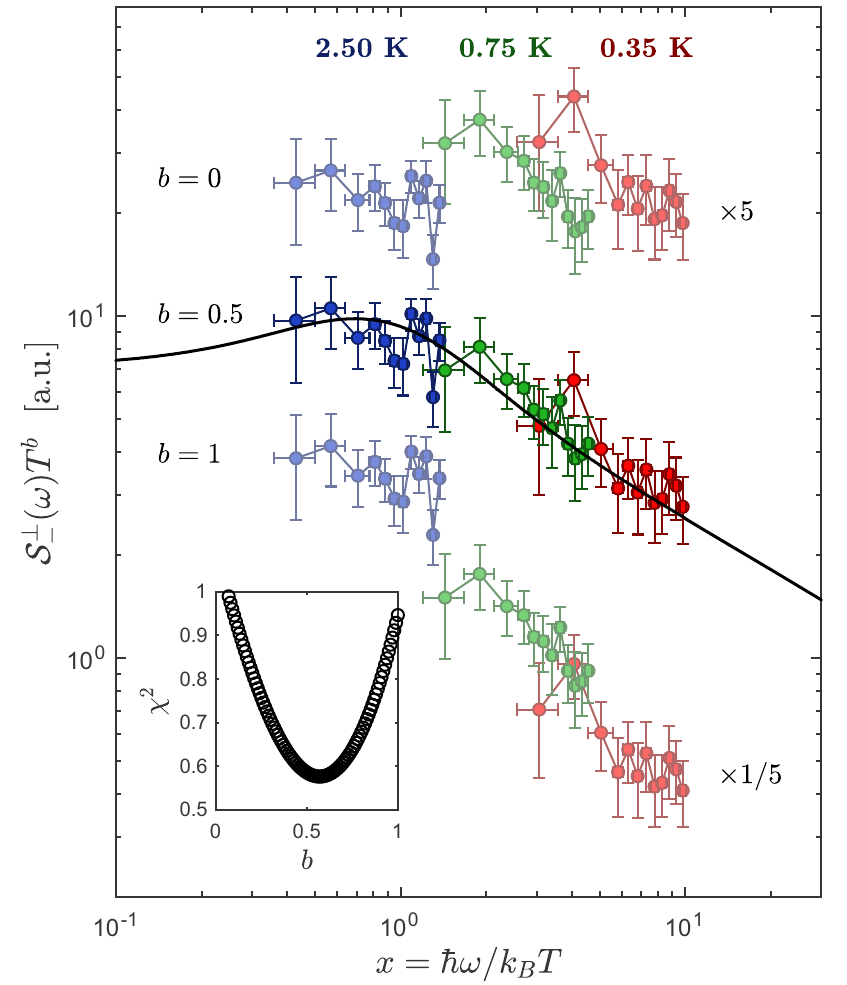}
\caption{\label{fig:StructureFactorScaling}Scaling plot of the antisymmetric transverse local dynamic structure factor component $\mathcal{S}_-^{\bot}(\omega)$  near the critical field at $H=6.5$~T. For the predicted scaling exponent $b=1/2$ all three data sets collapse onto a single continuous line. The inset shows the quality of the data collapse for different values of the critical exponent. Best scaling is found for $b=0.57(10)$. The solid line corresponds to the calculated scaling function.}
\end{figure}

Using this approach and integrating the data over $q_\|$, we obtain the local dynamic structure factor $\mathcal{S}_-^{\perp}(\omega)$. The results for three temperatures $T=0.35$, $0.75$, $2.5$~K obtained at $H=6.5$~T (almost exactly at $H_c$) are plotted in symbols in the scaling plot Fig.~\ref{fig:StructureFactorScaling}, using three different values of the scaling exponent $b$. The value of $b$ that produces the best data collapse was obtained as in the analysis of specific heat, using a 5\textsuperscript{th} degree polynomial fit. The resulting magnitude of the data mismatch $\chi^2$ is plotted in the inset of Fig.~\ref{fig:StructureFactorScaling}. From its minimum we determine $b=0.57(10)$ \footnote{The considerable experimental error on the estimate of the critical exponent $b=0.57(10)$ is due to the relatively small number of data points and the very limited overlap in $x$ between different data sets. As the theoretical value $b=1/2$ is within the error bar of the experimental estimate $b=0.57(10)$, within experimental precision we do not observe any deviations from scaling.},
where the theoretical value $b=1/2$ is within the error bar of the experimental estimate. Disregarding this difference we can consider the scaling plot with $b=1/2$ to be our experimental measurement of the scaling function $\Phi(x)$ in Eq.~\ref{phi}. 
Our present data cover both the range $\hbar\omega>k_BT$, which essentially reflects zero temperature properties of a dilute bose gas, and the quantum relaxation regime $\hbar\omega<k_BT$, where magnons strongly interact with thermally excited partners \cite{Sachdev1994}. In this our results qualitatively surpass the previous study on spin chains \cite{Blosser2017}. In that work, due the narrow energy range accessible, the scaling exponent could not be independently determined from experiment. Moreover, only the low temperature regime was covered, where measurements at several different temperatures do not contain additional information as they all probe the same $T=0$ dynamics.

The main result of the present work is that the measured scaling function is in excellent agreement with the exact result for this class of phase transitions (solid line in Fig.~\ref{fig:StructureFactorScaling}) that was evaluated numerically following Refs.~\cite{Korepin1990, Korepin1997Book, Blosser2017}.
The comparison to the arbitrarily normalized neutron scattering data was obtained by only fitting an overall scale factor (vertical shift along the logarithmic ordinate). This is different from previous studies of scaling of the dynamic structure factor in spin chains \cite{ Lake2005,Haelg2015_CuDCl} and ladders \cite{Povarov2015}, which focused on the $z=1$ Tomonaga-Luttinger liquid regime. For that situation the scaling function is also known exactly, but explicitly depends on the Luttinger parameter, which in turn depends on the applied magnetic field and the magnitude of XXY anisotropy in the system. In contrast, in the present case of $z=2$ quantum criticality, to apply the exact theoretical result, we did not require any information about BPCB other than it being one-dimensional, free of magnetic anisotropy and having a quadratic magnon dispersion relation at the transition.

Our work is entirely devoted to the universal low energy dynamics in the system. Nevertheless, we would like to draw the reader's attention to a peculiar non-universal spectral feature that emerges in a slightly magnetized ladder. Specifically, we observe a distinct splitting of both the middle and upper triplet bands at their maxima. 
Further work, particularly numerical simulations, will be required to understand this behavior. To this end, in the supplement we present some additional data highlighting this feature \cite{SM}. Here, we may only suggest that the appropriate language to describe it may be found in the mapping of the insulating spin ladder to a one-dimensional $t-J$ model as suggested in Ref. \cite{Bouillot2011}. Clearly, in-depth studies will be needed to clarify this issue.

In summary, long standing exact theoretical results for the universal finite-temperature scaling behavior of both specific heat and the local dynamic structure factor at the $z=2$ $d=1$ quantum critical point have been confirmed experimentally.

{\it Acknowledgement.}
We would like to thank Stanislaw Galeski and Severian Gvasaliya (ETHZ) as well as the sample environment team of the ISIS facility for their help with our experiments.  This work is partially supported by the Swiss National Science Foundation under Division II.

\bibliography{referencesBPCB}

\clearpage


\section*{Supplementary material}

# Supplemental material for "z = 2 quantum critical dynamics in a spin ladder"

D. Blosser,[1,*] V. K. Bhartiya,[1] D. J. Voneshen,[2] and A. Zheludev[1,†]

[1] *Laboratory for Solid State Physics, ETH Zürich, 8093 Zürich, Switzerland*
[2] *ISIS Facility, Rutherford Appleton Laboratory, Chilton, Didcot, Oxon OX11 0QX, United Kingdom*
(Dated: October 27, 2018)

### Determination of the critical field

Inelastic neutron scattering measurements of the magnon dispersion have been performed at various values of magnetic field. From these measurements the spin gap is determined. From a linear fit of the spin gap vs. applied magnetic field, we find $H_c = 6.66(6)$ T and $g = 2.16(3)$ in agreement with previously published estimates [1].

### Triplet dispersion and quasi-particle mass

For the strong rung ladder with $\gamma = J_{\mathrm{Leg}}/J_{\mathrm{Rung}} \ll 1$, a high order expansion in $\gamma$ yields an excellent approximation for the triplet dispersion relation. In Ref. 2 the following result is obtained up to third order in $\gamma$:

$$\epsilon(q_\parallel) = J_{\mathrm{Rung}} \left( 1 + \gamma \cos(q_\parallel) + \frac{\gamma^2}{4} \left( 3 - \cos(2q_\parallel) \right) - \frac{\gamma^3}{8} \left( 2\cos(q_\parallel) + 2\cos(2q_\parallel) - \cos(3q_\parallel) - 3 \right) \right). \tag{1}$$

Fitting this dispersion to our neutron scattering data, we find good agreement for $J_{\mathrm{Leg}}/k_B = 3.54(3)$ K and $J_{\mathrm{Rung}}/k_B = 12.67(6)$ K ($\gamma = 0.28$) as shown in Fig. 1. These estimates are fully consistent with previously published values [1]. In the vicinity of the parabolic minimum at $q_\parallel = \pi$, we thus obtain

$$\epsilon(q_\parallel) \approx \frac{1}{2} J_{\mathrm{Min}}(q_\parallel - \pi)^2 = \frac{\hbar^2(q_\parallel - \pi)^2}{2m}, \tag{2}$$

where $J_{\mathrm{Min}}/k_B = 5.07(10)$ K is the band curvature and $m = \hbar^2/J_{\mathrm{Min}}$ the triplet quasi-particle mass.

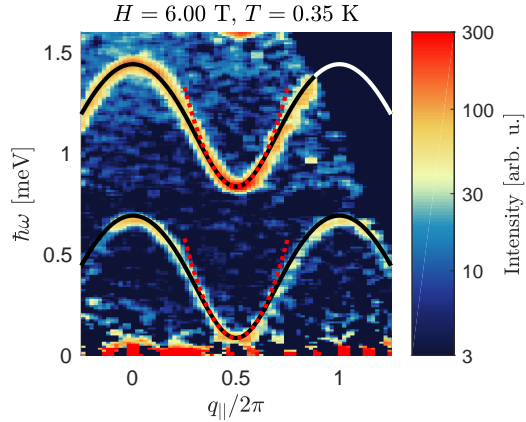

FIG. 1. False color plot of the background subtracted inelastic neutron scattering intensity measured at $H = 6$ T. For the middle triplet the black solid line corresponds to eqn. 1 with $J_{\mathrm{Leg}}/k_B = 3.54$ K and $J_{\mathrm{Rung}}/k_B = 12.67$ K. The red dotted line shows the parabolic approximation (eqn. 2) at the dispersion minimum. For the low energy triplet the same curves are simply shifted by the Zeeman energy.

## Separation of symmetric and anti-symmetric excitations

The neutron scattering intensity as measured in our experiments on the spin ladder compound BPCB is given by

$$\mathcal{I}(\mathbf{Q}, \omega) \propto |F(\mathbf{Q})|^2 \left[ 4 \left( 1 - \frac{Q_z^2}{Q^2} \right) \left[ s^+(\mathbf{Q}) \mathcal{S}_+^{zz}(q_\parallel, \omega) + s^-(\mathbf{Q}) \mathcal{S}_-^{zz}(q_\parallel, \omega) \right] \right. \tag{3}$$
$$\left. + \left( 1 + \frac{Q_z^2}{Q^2} \right) \left[ s^+(\mathbf{Q}) \mathcal{S}_+^{\perp}(q_\parallel, \omega) + s^-(\mathbf{Q}) \mathcal{S}_-^{\perp}(q_\parallel, \omega) \right] \right].$$

Here $F|\mathbf{Q}|^2$ is the magnetic form factor of the $Cu^{2+}$ ion [3, 4] and $q_\parallel = \mathbf{Q} \cdot \mathbf{a}$ is the wave vector transfer along the leg of the ladder. The rung structure factors read [5, 6]

$$s^\pm(\mathbf{Q}) = \frac{1}{4} \left( 2 \pm \cos(\mathbf{Q} \cdot \mathbf{d}_1) \pm \cos(\mathbf{Q} \cdot \mathbf{d}_2) \right), \tag{4}$$

where the vectors $\mathbf{d}_{1,2} = [0.3904, \pm 0.1598, 0.4842]$ in relative lattice units define the orientation of the ladder rungs for the two inequivalent ladders per unit cell in BPCB [7].

Near the gapped-gapless quantum critical point, the low energy universal critical fluctuations are descendants of the low energy triplet band i.e. they are due to the transverse antisymmetric structure factor component $\mathcal{S}_-^{\perp}(q_\parallel, \omega)$. The non-universal fluctuations present at low energy are due to scattering within the lower triplet band, conserve the number of triplons and therefore lie in the symmetric longitudinal structure factor $\mathcal{S}_+^{zz}$. The remaining structure factor components $\mathcal{S}_-^{zz}(q_\parallel, \omega)$ and $\mathcal{S}_+^{\perp}(q_\parallel, \omega)$ of Eq. 3 do not carry any spectral weight at low energies. Thus for our experiments, at low energies and after correction for the magnetic form factor, the scattering intensity is given by

$$\mathcal{I}(\mathbf{Q}, \omega) \propto 4 \left( 1 - \frac{Q_z^2}{Q^2} \right) s^+(\mathbf{Q}) \, \mathcal{S}_+^{zz}(q_\parallel, \omega) + \left( 1 + \frac{Q_z^2}{Q^2} \right) s^-(\mathbf{Q}) \, \mathcal{S}_-^{\perp}(q_\parallel, \omega). \tag{5}$$

The low energy universal and non-universal excitations might therefore be separated either by their different spin polarization or by their different symmetry. A polarized neutron experiment at this scale is technically infeasible today. However, in the present data sets we can separate symmetric and antisymmetric excitations as follows: At every fixed $(\tilde{q}_\parallel, \tilde{\omega})$, from the modulation of the measured scattering intensity along the non-dispersive reciprocal space directions $q_b, q_c$, we can deduce $\tilde{\mathcal{S}}_+^{zz} = \mathcal{S}_+^{zz}(\tilde{q}_\parallel, \tilde{\omega})$ and $\tilde{\mathcal{S}}_-^{\perp} = \mathcal{S}_-^{\perp}(\tilde{q}_\parallel, \tilde{\omega})$. In practice we proceed as follows:

First, at a fixed $(\tilde{q}_\parallel, \tilde{\omega})$ we integrate the measured scattering intensity $\mathcal{I}(\mathbf{Q}, \omega)$ over two regions where $s^+(\mathbf{Q}) > 0.7$ or $s^-(\mathbf{Q}) > 0.7$, respectively:

$$\mathcal{I}_1 = \iint_{s^+(\mathbf{Q}) > 0.7} \mathrm{d}q_b \, \mathrm{d}q_c \, \mathcal{I}(\mathbf{Q}, \tilde{\omega}), \qquad \mathcal{I}_2 = \iint_{s^-(\mathbf{Q}) > 0.7} \mathrm{d}q_b \, \mathrm{d}q_c \, \mathcal{I}(\mathbf{Q}, \tilde{\omega}), \qquad \mathbf{Q} = (\tilde{q}_\parallel, q_b, q_c). \tag{6}$$

These two intensities each contain scattering due to symmetric and antisymmetric excitations: (Since the scattering intensity in our experiment was not calibrated to absolute units, we ignore the proportionality relation of Eq. 5. The result will also be in arbitrary units.)

$$\mathcal{I}_1 = \alpha_1 \tilde{\mathcal{S}}_+^{zz} + \beta_1 \tilde{\mathcal{S}}_-^{\perp}, \qquad \mathcal{I}_2 = \alpha_2 \tilde{\mathcal{S}}_+^{zz} + \beta_2 \tilde{\mathcal{S}}_-^{\perp}. \tag{7}$$

The coefficients $\alpha_{1,2}$ and $\beta_{1,2}$ are calculated by integrating the prefactors of Eq. 5 in the same way as the measured scattering intensities. Now we can simply solve the two linear equations for the two structure factor components:

$$\tilde{\mathcal{S}}_+^{zz} = -\frac{\beta_2 \mathcal{I}_1 - \beta_1 \mathcal{I}_2}{\alpha_2 \beta_1 - \alpha_1 \beta_2}, \qquad \tilde{\mathcal{S}}_-^{\perp} = -\frac{-\alpha_2 \mathcal{I}_1 + \alpha_1 \mathcal{I}_2}{\alpha_2 \beta_1 - \alpha_1 \beta_2}. \tag{8}$$

Naturally, in a scattering experiment, the measured intensity is associated with a statistical error. These errors for the integrated intensities $\mathcal{I}_{1,2}$ we denote $\sigma_{1,2}$. For the extracted ladder structure factors $\tilde{\mathcal{S}}_+^{zz}$ and $\tilde{\mathcal{S}}_-^{\perp}$ we find error estimates

$$\sigma_+ = \sqrt{\frac{\beta_2^2 \sigma_1^2 + \beta_1^2 \sigma_2^2}{(\alpha_2 \beta_1 - \alpha_1 \beta_2)^2}}, \qquad \sigma_- = \sqrt{\frac{\alpha_2^2 \sigma_1^2 + \alpha_1^2 \sigma_2^2}{(\alpha_2 \beta_1 - \alpha_1 \beta_2)^2}}. \tag{9}$$

Of course when defining the distinct regions in reciprocal space by $s^+(\mathbf{Q}) > 0.7$ or $s^-(\mathbf{Q}) > 0.7$, respectively, the threshold value of 0.7 is arbitrary. In fact we have obtained the same results with different choices of this value.

In summary, repeating the above procedure at every point $(q_\parallel, \omega)$ at low energies, we obtain separately the symmetric longitudinal and antisymmetric transverse ladder structure factor components $\mathcal{S}^{zz}_+(q_\parallel, \omega)$ and $\mathcal{S}^\perp_-(q_\parallel, \omega)$, respectively. The plots in Fig. 3(h,i,j) of the main paper illustrate this decomposition for the neutron spectrum obtained at $H = 6.75$ T and $T = 0.35$ K.

Finally, we comment on the applicability of this procedure: For a simple compound with a single two-leg ladder per unit cell, this separation of structure factor components works perfectly. In the studied compound BPCB there are two inequivalent ladders per unit cell. However, their rung-vectors differ along one axis ($\mathbf{d}_{1,2} = [0.3904, \pm 0.1598, 0.4842]$ relative lattice units). Thus in large portions of reciprocal space the rung-structure factors $s^\pm(\mathbf{Q})$ take very different values and our procedure works very well. However, in an even more complex crystal structure with several inequivalent ladders per unit cell, the two rung-structure factors $s^\pm(\mathbf{Q})$ might not vary much and take very similar values. In this case there would be very little 'contrast' between the symmetric and antisymmetric excitations and separating the two components might be difficult or impossible.

### Evolution of the high energy triplet branches upon magnetizing the ladder

The present work has exclusively focused on the low energy universal fluctuations near the $z = 2$ quantum critical point. However, upon magnetizing the ladder, we also observe a peculiar splitting of the two high energy triplet bands. In this section we present our data in a slightly different fashion than in the main paper so as to highlight these features.

Our experiment on the LET spectrometer made use of repetition rate multiplication [8]. This means, we have simultaneously obtained data with different incident energy neutrons: $E_i = 1.35, 2.20, 4.20, 10.97$ meV. The energy resolution (at zero energy transfer) achieved in these data sets is $\Gamma = 0.019, 0.036, 0.097, 0.409$ meV (full width at half maximum), respectively. In the $E_i = 1.35$ meV channel the intensity was very low. Thus, in the main paper we use the $E_i = 2.20$ meV data for studying the low energy spin dynamics. In the following, we also present data obtained at $E_i = 4.20$ meV. At the cost of lower resolution, this data extends to higher energy transfer, also covering the high energy triplet mode.

In Fig. 2 we summarize all data obtained in the vicinity of the critical field $H_c = 6.66(6)$ T. In this figure, the $E_i = 2.20$ meV data is plotted exactly as in the main paper, except for the addition of one data set obtained at $H = 6.00$ T, $T = 0.35$ K. Upon magnetizing the ladder we observe a split-off feature appearing near the band maxima of both the middle and high energy triplet bands. To quantify this evolution, in figure 3 we show cuts through the dispersion maximum at $q_\parallel/2\pi = 0$.

We make the following observations:

- Upon magnetizing the ladder, near the band maxima of the middle and the upper triplet band, a new feature appears and gains spectral weight.

- This branch does not split off the main magnon band. It appears and gains spectral weight with the 'gap' between the new feature and the main triplet band roughly constant at $\Delta \sim 0.15$ meV.

- The band maximum of the low energy triplet remains essentially unchanged upon slightly magnetizing the ladder.

Finally we note that in all data sets obtained at $E_i = 4.2$ meV, there is an apparent 'line' of low intensity at $\hbar\omega \approx 1.9$ meV, independent of magnetic field or temperature. Most likely, this is an experimental artifact and not due to spin dynamics in the sample.

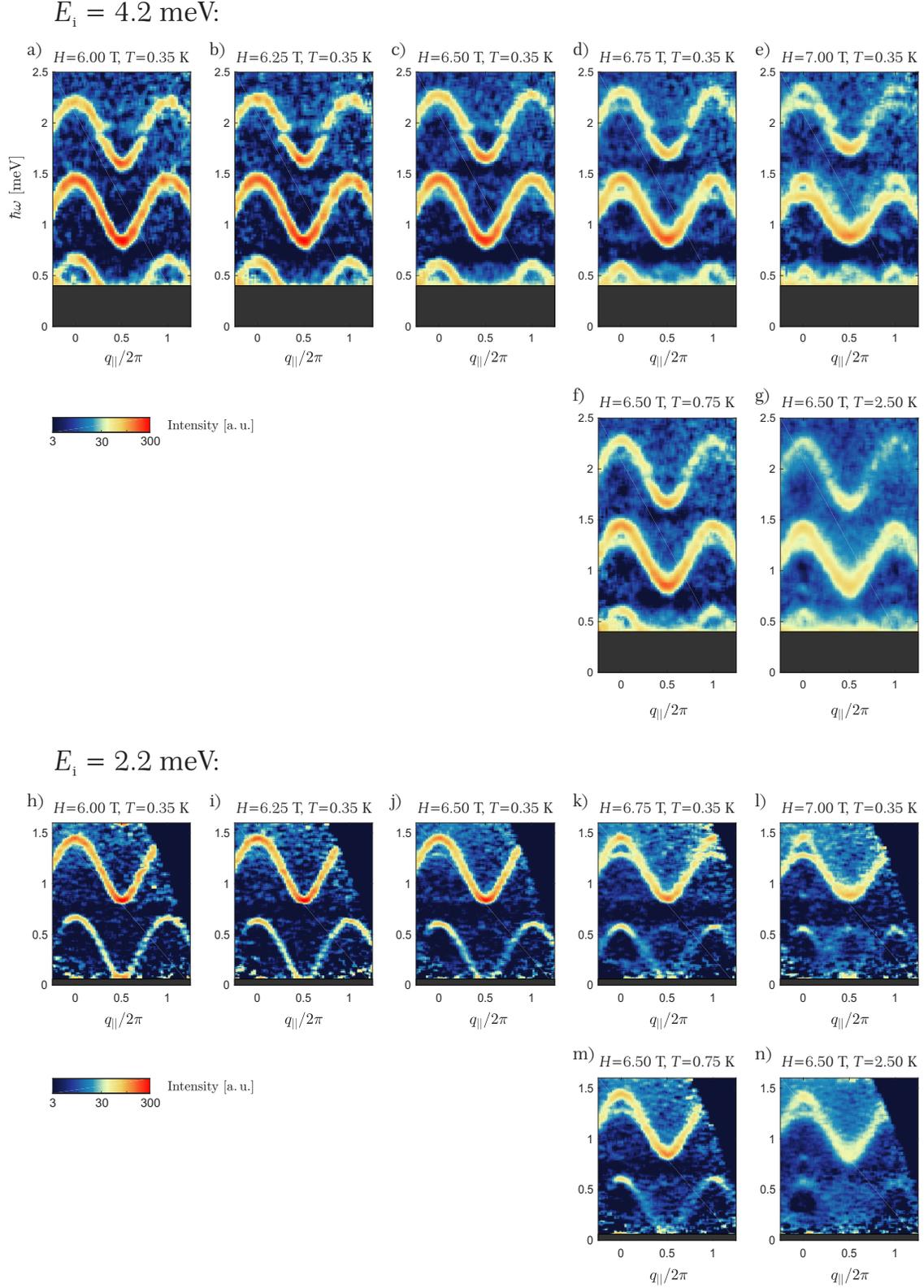

FIG. 2. Inelastic neutron scattering spectra measured with $E_i = 4.2$ meV (a-g) and $E_i = 2.2$ meV (h-n) incident energy neutrons, respectively. The data sets are integrated fully along the two non-dispersive reciprocal space directions. Plots (i-n) are identical to the ones presented in Fig. 3 of the main paper.

Upon magnetizing the ladder we observe a distinct splitting of the two high energy triplet bands at their maxima.

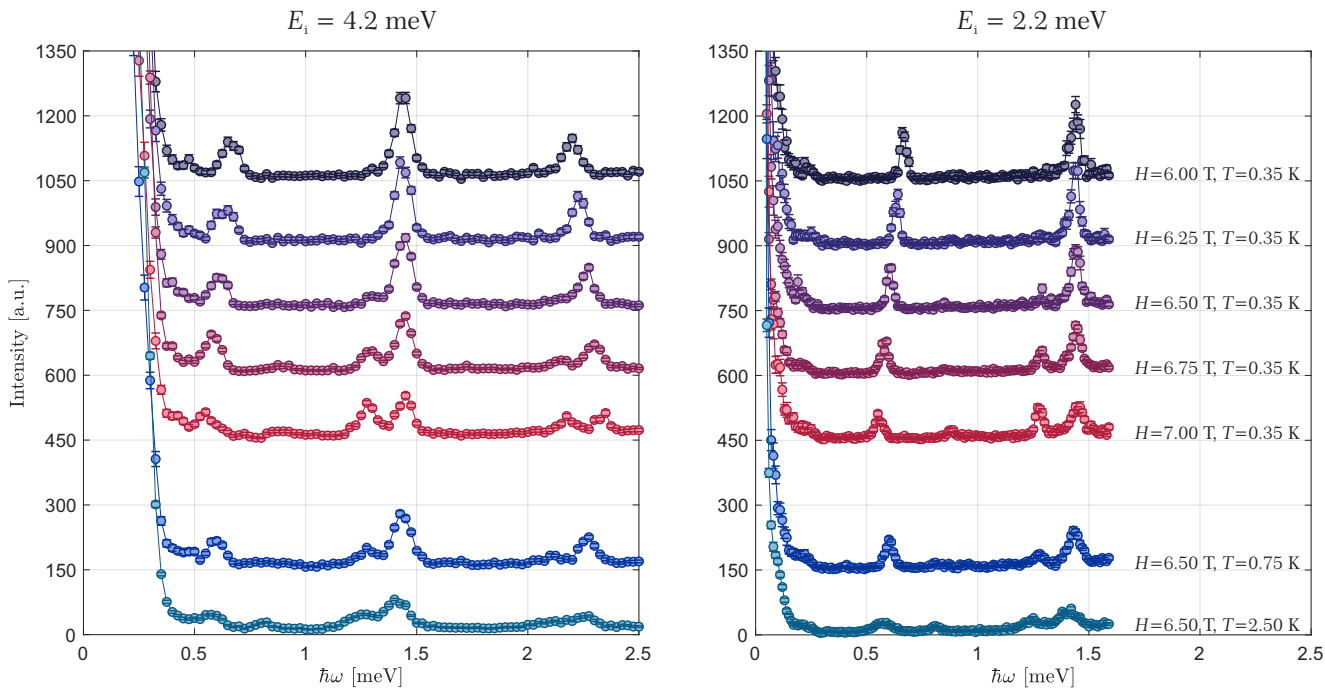

FIG. 3. Cuts through the dispersion maximum: The measured scattering intensity is integrated in a narrow slice $q_{\parallel}/2\pi \in [-0.05, 0.05]$ and fully along the two non-dispersive reciprocal space directions.

In an increasing magnetic field we observe the Zeeman-splitting of the tripplet bands. Upon magnetizing the ladder ($H_c = 6.66(6)$ T), both for the middle and the high energy triplet band, we observe the 'split-off' feature appearing as a new peak gaining intensity.


\clearpage

%
%
%

\end{document}